\documentclass[a4paper,fleqn,usenatbib]{mnras}

\usepackage{newtxtext,newtxmath}

\usepackage[T1]{fontenc}
\usepackage{ae,aecompl}

\usepackage{graphicx}	
\usepackage{amsmath}	
\usepackage{amssymb}	
\usepackage[multi-part-units=single]{siunitx}
\usepackage{mathtools}
\usepackage{xcolor}
\usepackage{cleveref}
\usepackage{commath}
\usepackage{subfig}

\makeatletter
\providecommand\add@text{}
\newcommand\sieq[1]{%
	\gdef\add@text{\textcolor{lightgray}{[\si{#1}]}\gdef\add@text{}}}%
\renewcommand\tagform@[1]{%
	\maketag@@@{\llap{\add@text\quad}(\ignorespaces#1\unskip\@@italiccorr)}%
}
\makeatother

\DeclareSIUnit \parsec {pc}

\title[A sceptical analysis of Quantized Inertia]{A sceptical analysis of Quantized Inertia}

\author[M. Renda]{
Michele Renda,$^{1}$\thanks{E-mail: michele.renda@cern.ch } \\
$^{1}$ Departament of Elementary Particle Physics, IFIN-HH, Reactorului 30, P.O.B. MG-6, 077125, M\u{a}gurele, Romania\\
}

\date{Accepted XXX. Received YYY; in original form ZZZ}

\pubyear{2019}

\begin{document}
\label{firstpage}
\pagerange{\pageref{firstpage}--\pageref{lastpage}}
\maketitle

\begin{abstract}

We perform an analysis of the derivation of Quantized Inertia (QI) theory, formerly known with the acronym MiHsC, as presented by \citet{mcculloch_modelling_2007,mcculloch_inertia_2013}. Two major flaws were found in the original derivation. We derive a discrete black-body radiation spectrum, deriving a different formulation for $F(a)$ than the one presented in the original theory. We present a numerical result of the new solution which is compared against the original prediction.  
\end{abstract}




\section{Introduction} \label{sec:into}

The discrepancy between observed galaxies rotation curves and the prediction using the known laws of orbital kinematics was initially observed by \citet{rubin_rotational_1980}, and it is now an accepted phenomenon.

Several theories were developed to justify such discrepancies, such as the 
existence of a dark matter halo \citep{rubin_rotation_1983}, or the 
existence of a Modified Newtonian Dynamics, MoND 
\citep{milgrom_modification_1983} at galactic scales.

As today, no direct evidence of dark matter was detected, though many experiments such as XENON100 \citep{aprile_xenon100_2012} and SuperCDMS \citep{supercdms_collaboration_search_2014} are looking for signal candidates. Some models support the idea that dark matter particles could be created at LHC and efforts in this direction are in progress \citep{abercrombie_dark_2015, mitsou_overview_2015, liu_enhancing_2019}.

MoND models remove the necessity for dark matter candidates introducing a modified law of motion for low accelerations:
\begin{align}
F &= \begin{dcases*}
m \, a  & when $a \gg a_0 $ \\
m \, \dfrac{a^2}{a_0} & when $a \ll a_0 $
\end{dcases*} \sieq{\newton}
\end{align}

This approach has been criticized due to the requirement of an arbitrary parameter $a_0$ and because it does not predict the dynamics of galaxy clusters \citep{aguirre_problems_2001, sanders_clusters_2003}. A new theory, by McCulloch, proposes a solution to the discrepancies observed in the galaxies' rotation curves. This theory, named Modification of inertia resulting from a Hubble-scale Casimir effect (MiHsC) or Quantized Inertia (QI), may give a model for the galaxies' rotation curves \citep{mcculloch_testing_2012} and explain some other phenomena like the Pioneer anomaly \citep{mcculloch_modelling_2007}, the flyby anomalies \cite{mcculloch_modelling_2008}, the Em-drive 
\citep{mcculloch_testing_2015}, opening the way for propellant-less spacecraft propulsion \citep{mcculloch_propellantless_2018}. In addition, this theory provides also an intuitive explanation for objects' inertia \citep{mcculloch_inertia_2013}.

The main strong points of this theory, as shown by \cref{eq:formula_qi}, are the absence of arbitrary tunable parameters (being based on universal constants like the Hubble constant and the speed of light), its simple formulation and the wide range of phenomena it seems to explain. Its main weak point is the fact it assumes the existence of the Unruh radiation \citep{unruh_notes_1976}, which is still not experimentally measured in nature, although some recent simulations seems to confirm its existence \citep{hu_quantum_2019}.

Quantized Inertia has collected some criticisms by mainstream press \citep{koberlein_quantized_2017}, but, as today, no critical analysis was published in a peer-reviewed journal on this subject. 

In the next sections we will present our analysis of Quantized Inertia: in \cref{sec:recap} we will perform a brief recapitulation of the theory as presented by \cite{mcculloch_modelling_2007,mcculloch_inertia_2013}, in \cref{sec:correction1,sec:correction2} we will present two major flaws we found in its derivation and we propose some corrections and finally, in \cref{sec:conclusion}, we will present our considerations about the validity of the whole theory.  

\section{Recapitulation on Quantized Inertia}  \label{sec:recap}

Quantized Inertia states two important affirmation:

\begin{enumerate}
	\item There exists a minimum acceleration any object can ever have: $a_0 = 2 c^2 /\Theta = \SI{2e-10}{\meter\per\second\squared}$  \citep[see][sect. 2]{mcculloch_low_2017}. Below such a value, the object's inertia becomes zero causing the object's acceleration to increase to the minimum value.
	\item Inertia is caused by the Unruh radiation imbalance between the cosmic and the Rindler horizon \citep[fig.1]{mcculloch_inertia_2013}.
\end{enumerate}

The rationale behind the first point is this: according to the Unruh radiation law \citep{unruh_notes_1976}, every accelerating object will feel a background temperature:
\begin{align}
T &= \cfrac{\hbar a}{2 \pi c k} \sieq{\kelvin} \label{eq:unruh}
\end{align}
where $\hbar$ is the reduced Planck constant, $c$ the speed of light in vacuum, $k$ the Boltzmann constant and $a$ the object's acceleration. It is important to notice that this temperature is very tiny: for an object acceleration of \SI{1}{\meter\per\second\squared}, the temperature will be $T \approx \SI{4e-21}{\kelvin}$, making very difficult any experimental detection.

Planck's law states such background will emit black-body radiation with spectrum:
\begin{align}
b_{\lambda}(\lambda,T) &= \dfrac{2 h c^2}{\lambda^5}\dfrac{1}{e^{h c/\lambda k T} - 1} \sieq{\watt\per\meter\per\steradian\per\square\meter} \label{eq:planck_wave_power_radiance}
\intertext{with a peak wavelength:}
\lambda_{\text{peak}} &=  \dfrac{h c}{a_5 k T} \sieq{\meter} \label{eq:planck_wave_peak}
\end{align} 
where $a_5\approx \SI{4.96511423174}{}$ is the solution of the transcendental equation $5(1 - e^{-x}) = x$ and $h$ is Planck's constant. We would like to remark that the Planck's law describes an \emph{unconstrained} system at equilibrium. This is the case in a classic black-body radiation experiment where the radiation wavelength is much smaller than the cavity size, leading to a continuous spectrum.

However, for very low temperatures, the associated radiation wavelength may become bigger than the cosmic horizon, defined as the sphere with radius equal to the Hubble distance. We introduce so the Hubble diameter defined as:
\begin{align}
\Theta = \dfrac{2 c}{H_0} \approx \SI{2.607e26}{\meter} \label{eq:theta}
\end{align}
where $H_0$ is the Hubble constant\footnote{In this paper we assume $H_0 = \SI[separate-uncertainty=true]{2.3 \pm 0.9e-18}{\per\second}$, as used by \citet[sect. 2.1]{mcculloch_modelling_2007}, for easier results comparison with the original papers.} and $c$ is the speed of light in vacuum. In this case we can not consider the spectra continuous any more, because, by principle, only the wavelengths fitting twice the Hubble diameter can ever exist:
\begin{align}
\lambda_n = \frac{2 \Theta}{n} \quad \text{where} \quad n=1,2,3\ldots\infty \sieq{\meter}
\end{align}

\begin{figure}
	\centering
	\includegraphics[width=\columnwidth]{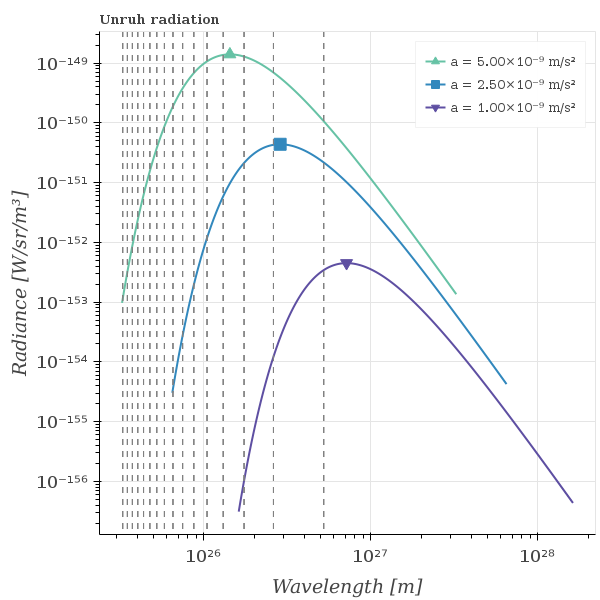}
	\caption{Plot of the Unruh radiation spectra for different object accelerations. The marks represent the $\lambda_{\text{peak}}$ for the given acceleration. The last dashed line on the right represents the biggest wavelength fitting inside twice the Hubble diameter.}
	\label{img:radiance}
\end{figure}

The minimum acceleration arises from two phenomena: for lower accelerations, the object experiences lower Unruh background radiation due to a) the shift of the spectra behind the Hubble diameter and b) a more sparse sampling of the black body radiation spectra, as shown in \cref{img:radiance}.

The consideration expressed above were used to define a Hubble Scale Casimir-like effect: using a not better specified \emph{direct calculation},  \citet[sect. 2.2]{mcculloch_modelling_2007} affirms there is a linear relation between the continuous and the discrete sampling of the Unruh radiation spectra. The ratio between the two sampling, denoted with $F(a)$, is considered to be a linear function. Assuming that for $\lambda_{\text{peak}} \rightarrow 0$ we have the classical case ($F=1$), and for $\lambda_{\text{peak}} \rightarrow 4 \Theta$ no Unruh radiation is sampled ($F=0$), this relation was proposed by \citet{mcculloch_modelling_2007}:
\begin{align}
m_I = F(a) \; m_i = \left(1 - \cfrac{\beta \pi^2c^2}{a \Theta}\right) m_i \sieq{\kilogram} \label{eq:formula_qi}
\end{align}
where $\beta = 1 / a_5 \approx \SI{0.2}{}$, $a$ is the acceleration modulus, $m_i$ is the classic inertial mass and $m_I$ the modified inertial mass.

\section{Correction 1} \label{sec:correction1}
We focus our attention on the derivation of \cref{eq:formula_qi}, based on the linear relation:
\begin{align}
m_I = F(a) \; m_i = \cfrac{B_s(a)}{B(a)} \; m_i \sieq{\kilogram} \label{eq:f_factor}
\end{align}
where $B_s$ is the sampled (discrete) black body radiance and $B$ is the classical one. The value of $B$ can be found integrating \cref{eq:planck_wave_power_radiance}: 
\begin{align}
B(T) &= \int_{0}^{\infty} b_{\lambda}(\lambda,T) \; d\lambda = \dfrac{2 \pi^4 k^4}{15 h^3 c^2} T^4 \sieq{\watt\per\meter\squared\per\steradian}\label{eq:planck_int}
\end{align}
while the determination of $B_s$ is more complex and will be discussed in \cref{sec:planck_contrained}.

\subsection{Derivation of Planck's law for unconstrained cavities} \label{sec:planck_uncontrained}
If we have a cubic cavity with side $L$, we can have an infinite number of independent radiation modes. Each mode can be defined by three non-negative integers, $l,m,n$, such that the wave fits entirely in twice the cavity side:
\begin{align}
\lambda_x = \cfrac{2 L}{l} \qquad \lambda_y = \cfrac{2 L}{m} \qquad \lambda_z = \cfrac{2 L}{n} \sieq{\meter} \label{eq:rel_lambda_lmn}
\end{align}
Using this notation, it is possible to define a new quantity named wave-vector defined as:
\begin{align}
k_x &= \cfrac{2 \pi} {\lambda_x} = \cfrac{\pi l}{L} \sieq{\per\meter} \quad k_y = \cfrac{2 \pi} {\lambda_y} = \cfrac{\pi m}{L} \quad k_x = \cfrac{2 \pi} {\lambda_z} = \cfrac{\pi n}{L} \sieq{\per\meter} \sieq{\per\meter} \label{eq:rel_lambda_k}
\end{align}
so any wave can be expressed as:
\begin{align}
A(\textbf{r}, t) &= A_0 \; sin(\textbf{k} \cdot \textbf{r} - \omega t)
\end{align}

\begin{figure}
	\centering
	\subfloat[Continuous integral\label{img:wavemodes_continuos}]{
		\includegraphics[width=\columnwidth]{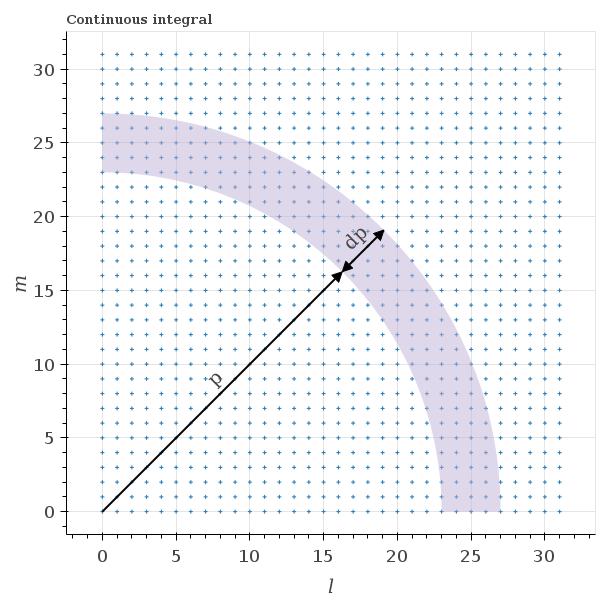}
	}
	\hfill
	\subfloat[Discrete sum\label{img:wavemodes_discrete}]{%
		\includegraphics[width=\columnwidth]{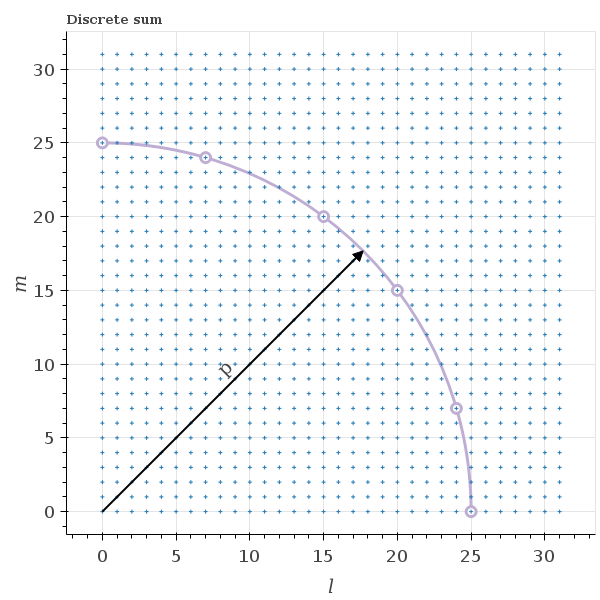}
	}
	\caption{Plane section of the $l,m,n$ volume: each cross represents a wave-mode. Points near to the origin will have longer wavelengths while points with the same modulus will share the same wavelength and energy.}
	\label{img:wavemodes}
\end{figure}

Using this formalism,  we can express each wave in a cavity using three non-negative integers, $l,m,n$: smaller integers represent longer wavelengths, while higher values represent shorter ones. We can represent these points in a graph, as shown in \cref{img:wavemodes}.

Every point represents a wave-mode in the cavity: the points with the same modulo will have the same energy, or, more concisely, if we define $p^2 = l^2 + m^2 + n^2$, for the same value of $p$, we have the same energy. The relation between $\lambda$ and $p$ now becomes:
\begin{align}
\lambda = \cfrac{2L}{p} \sieq{\meter} \label{eq:rel_lambda_p}
\end{align}

If we want to calculate the energy density, we have to sum the number of wave-modes with the same energy multiplied by their average energy and divide by volume:
\begin{align}
U(T) = \sum_{p} u(p,T)= \sum_{p} \cfrac{2 \; N(p) \; \overline{E}(p,T) \sieq{\joule\per\meter\cubed}}{L^3} \label{eq:def_U_p}
\end{align}
where $N(p)$ is the number of independent modes with wave-mode $p$,  $\overline{E}(p)$ is the average energy of that mode and the factor \SI{2}{} reflects the fact that each wave can have two independent polarizations. $\overline{E}(p)$ can be found using the Boltzmann distribution:
\begin{align}
\overline{E}(p,T) = E_p \; \dfrac{e^{-\frac{E_p}{kT}}}{\sum\limits_{p^2 = 1}^{\infty}{e^{-\frac{E_p}{kT}}}} = \frac{h c}{\lambda_p} \; \frac{1}{e^{h c / \lambda_p k T} - 1}  \sieq{\joule}
\end{align}
where $\lambda_p = 2 L / p $ and $E_p = h c / \lambda_p$.

For the determination of $N(p)$, we need to estimate the number of wave-modes for a given energy: for an unconstrained system, when $\lambda_{\text{peak}} \ll L$, we can suppose the wave-modes are so dense we can estimate them as the volume of a shell of a sphere with radius $p$ and thickness $\dif p$  (as shown in \cref{img:wavemodes_continuos}):
\begin{align}
\overline{N}(p) \; \dif p = \frac{1}{8} \; 4 \pi \; p^2 \dif p
\end{align}
where the factor $1/8$ reflects the fact we are only counting one octant ($l,m,n > 0$). We can now transform \cref{eq:def_U_p} into a \emph{continuous} sum, by frequency ($\nu = c p / 2 L $) or by wavelength ($\lambda = 2L / p$):
\begin{align}
U(T) = \int u(\nu,T) \dif{\nu} =  \int u(\lambda,T) \dif{\lambda} \sieq{\joule\per\meter\cubed} \label{eq:def_U}
\end{align}
where
\begin{align}
u(\nu,T) &= \cfrac{8 \pi h \nu^2}{c^3} \; \cfrac{1}{e^{h \nu / k T} - 1} \sieq{\joule\per\hertz\per\meter\cubed} \label{eq:def_u_n} \\
u(\lambda,T) &= \cfrac{8 \pi h c}{\lambda^5} \; \cfrac{1}{e^{h c / \lambda k T} - 1} \sieq{\joule\per\meter\per\meter\cubed} \label{eq:def_u_l}
\end{align}

\Cref{eq:def_u_n,eq:def_u_l} are often presented in the form of power radiance, which can be found multiplying them by $c/4\pi$:
\begin{align}
b(\nu,T) &= \cfrac{2 h \nu^3}{c^2} \; \cfrac{1}{e^{h \nu / k T} - 1} \sieq{\watt\per\hertz\per\steradian\per\square\meter} \label{eq:def_b_n} \\
b(\lambda,T) &= \cfrac{2 h c^2}{\lambda^5} \; \cfrac{1}{e^{h c / \lambda k T} - 1} \sieq{\watt\per\meter\per\steradian\per\square\meter} \label{eq:def_b_l}
\end{align}

\subsection{Derivation of Planck's law for constrained cavities} \label{sec:planck_contrained}
In \cref{sec:planck_uncontrained} we discussed the derivation of Planck's law because now we will use the same principle to derive a similar equation for constrained cavities, where the wavelength sizes are comparable to the cavity dimensions. This time we can not transform \cref{eq:def_U_p} in a continuous sum, but we have to handle it as an infinite \emph{discrete} sum.

The value of $E(p,T)$ can be found using the Boltzmann distribution:
\begin{align}
\overline{E}(p,T) = \overline{E}_{\lambda_p=2L/p}(\lambda_p, T) = \cfrac{h c}{\lambda_p} \; \cfrac{1}{e^{h c / \lambda_p k T} - 1} \sieq{\joule}
\end{align}
while the value $N_p$ are the number of modes where $l^2 + m^2 + n^2 = p^2$, as shown in \cref{img:wavemodes_discrete}. Unfortunately, this value can not be calculated analytically but, if we define $n=p^2$ we can find the value of $N(p)$ in the sequence \href{https://oeis.org/A002102}{A002102} \citep{sloane_encyclopedia_1995}. Using this definition, \cref{eq:def_u_l,eq:def_b_l} become, respectively:
\begin{align}
u_s(p,T) &= \dfrac{2 N(p)}{L^3} \;  \dfrac{hc}{\lambda_p} \; \dfrac{1}{e^{h c / \lambda_p k T}-1} \sieq{\joule\per\meter\cubed} \\
b_s(p,T) &= \dfrac{2 N(p)}{L^3} \;  \dfrac{hc^2}{4 \pi \lambda_p} \; \dfrac{1}{e^{h c / \lambda_p k T}-1} \sieq{\watt \per\steradian\per\meter\squared}
\end{align}
where $\lambda_p = 2L / p$. Finally, we can find the sum for all the modes as:
\begin{align}
U_s = \sum_{p^2 = 1}^{\infty} u_s(p, T) \sieq{\joule\per\meter\cubed} \\
B_s = \sum_{p^2 = 1}^{\infty} b_s(p, T) \sieq{\watt \per\steradian\per\meter\squared}
\end{align}

\subsection{Ratio between $B_s$ and $B$}

Using the results from the previous section and \cref{eq:f_factor,eq:theta,eq:rel_lambda_p}, now it is possible to find a new expression for the function $F(T)$ :
\begin{align}
F(T) = \cfrac{15 H^4_0 h^4}{128 \pi^5 k^4 T^4} \; \sum_{p^2 = 1}^{\infty} N(p) \; \dfrac{p}{e^{h p / 4 H_0 k T}-1} \label{eq:f_factor_T}
\end{align}
which can be expressed, using \cref{eq:unruh}, as a function of the object's acceleration:
\begin{align}
F(a) = \cfrac{30 \pi^3 H^4_0 c^4}{a^4} \; \sum_{p^2 = 1}^{\infty} N(p) \; \dfrac{p}{e^{ p \pi^2 c H_0 / a}-1} \label{eq:f_factor_a}
\end{align}

\begin{figure}
	\includegraphics[width=\columnwidth]{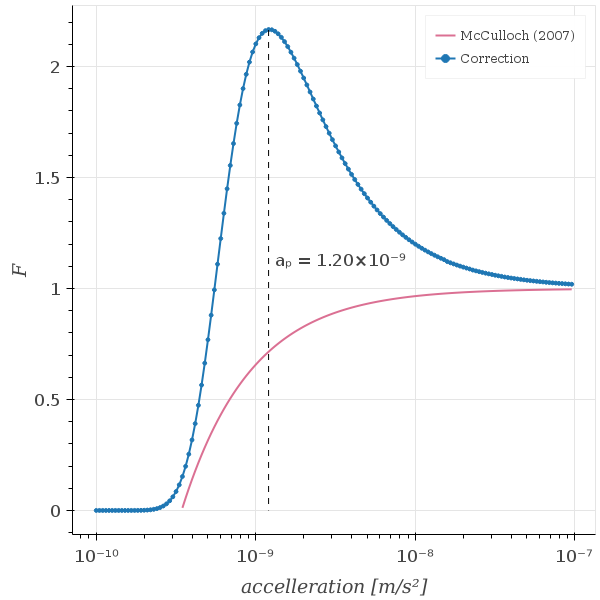}
	\caption{Plot of the $F(a) = B_s(a) / B(a)$ function for low accelerations. We can observe that $F(a \rightarrow \infty) = 1$ (classical case) and  $F(a \rightarrow 0) = 0$ and a peak around $a_p \approx \SI{1.20e-9}{\meter\per\second\squared}$. The marks represent the values in which the calculation was performed.}
	\label{img:f_factor}
\end{figure}

The term $N(p)$ make very difficult any analytical solution of \cref{eq:f_factor_a}, but it is possible to solve numerically, as shown in \cref{img:f_factor}. We can observe it is different from \cref{eq:formula_qi}: while we can observe that for $a > \SI{1e-8}{\meter\per\second\squared}$, $F = 1$ (classical case), and for $a < a_0$, $F=0$, as predicted by \cite{mcculloch_modelling_2007}, but we also have a critical point at $a_p \approx \SI{1.20e-9}{\meter\per\second\squared}$ where we have a maximum value for $F \approx \SI{2.17}{}$.

This point would represent a stable point because, if we apply a small force to an object around this critical value, the shape of $F(a)$ will stabilize the object's acceleration. At the knowledge of the authors, no such behaviour was ever measured or predicted theoretically by other models.

\section{Correction 2} \label{sec:correction2}
\begin{figure}
	\centering
	\includegraphics[width=0.80\columnwidth]{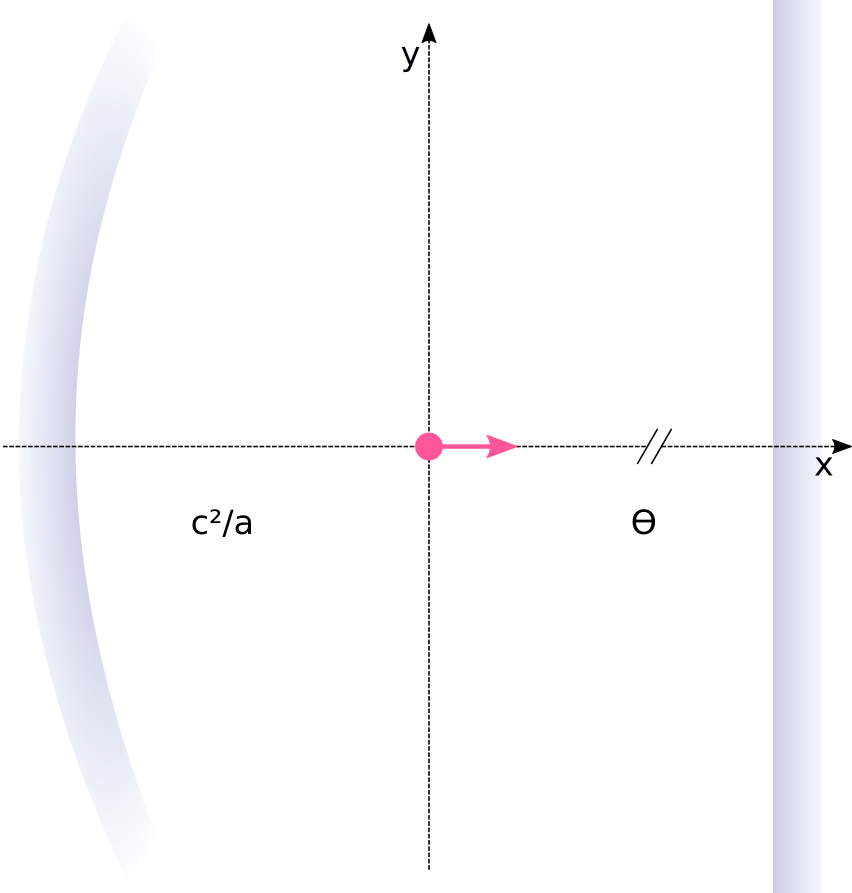}
	\caption{Schematic representation of the Cosmic and Rindler horizon as presented by \citet{mcculloch_inertia_2013}. If the object is accelerating to the right a Rindler horizon is formed on the left, disallowing some Unruh waves on that side and, consequentially, a lower radiation pressure. The radiation pressure imbalance will produce a force against the direction of acceleration.}
	\label{img:horizon}
\end{figure}

In this section, we will discuss the radiation imbalance between the Cosmic and Rindler horizon. In \citet{mcculloch_inertia_2013}, it is shown how applying \cref{eq:formula_qi} to an object moving along the direction $x$, as shown in \cref{img:horizon}, there will be an imbalance between the radiation pressure on the right, limited by the cosmic horizon, and the radiation pressure on the left, limited by the nearer Rindler horizon.

This radiation pressure imbalance will cause a force reacting against any acceleration similar in behaviour to the classical inertia.  It is shown in \cite{mcculloch_inertia_2013} (and partially corrected by \citet{gine_inertial_2016}), it is possible to express the force $\textbf{F}$ as: 
\begin{align}
\textbf{F} = - \cfrac{\pi^2 h A}{48 c V} \; \textbf{a} \sieq{\newton}
\end{align}
where $A$ is the object's radiation cross-section, smaller than the physical cross-section, $V$ is the object's volume and $a$ the modulus of the object's acceleration.

It is also shown that, if we assume the particle a cube with size equal to the Planck's length, $l_P = \SI{1.616e-35}{\meter}$, the model predicts an inertial mass of \SI{2.799e-8}{\kilogram}, which is \SI{29}{\percent} greater
than the Planck's mass, $m_P = \SI{2.176e-8}{\kilogram}$ \citep{gine_inertial_2016}.

Our main concern is how the energy density substitution was performed in  both \citet[eq. 9-10]{mcculloch_inertia_2013} and \citet[eq. 7-8]{gine_inertial_2016}:
\begin{align}
u = \cfrac{E}{V} = \cfrac{hc}{\lambda V}  \sieq{\joule\per\meter\cubed} \label{eq:rel_u_EV}
\end{align}

In this substitution, the authors imply that only the peak wavelength of the Unruh spectrum contributes to the energy densities. In reality, this is deeply incorrect because for classical accelerations (i.e. $a > \SI{1e-8}{\meter\per\second\squared}$), the peak wavelength contributes only a tiny part of the overall energy density. We can define a new quantity $G$, expressing the contribution of the peak wavelength to the radiation energy density as:
\begin{align}
G(a) &= \cfrac{u_s (p', a)}{U_s(a)} = \frac{N(p') \; \frac{p'}{e^{  p' \pi^2 c H_0 / a}-1}}{
	\sum\limits_{p^2 = 1}^{\infty} N(p) \;  \frac{p}{e^{ p \pi^2 c H_0 / a}-1}} 
\end{align}
where $p'$ is the wave-mode nearest to the peak wavelength, which can be calculated as:
\begin{align}
p' &= \sqrt{\text{round}\left(\frac{ a_5 \; a}{\pi^2 c H_0}\right)^2}
\end{align}

\begin{figure}
	\centering
	\includegraphics[width=\columnwidth]{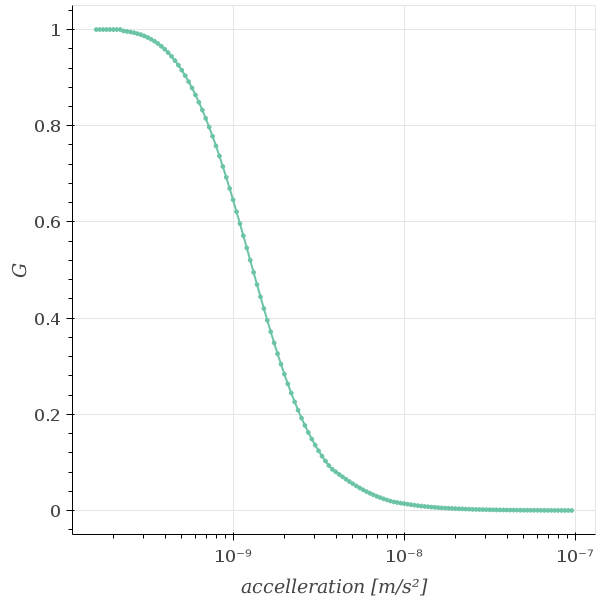}
	\caption{Plot of $G(a)$: it shows the contribution of the highest $u_s$ over the entire spectrum. For lower accelerations this value is near to one because only a few modes are allowed but for higher accelerations its contribution tends to zero.}
	\label{img:g_factor}
\end{figure}

In figure \cref{img:g_factor}, we can see the plot of $G(a)$ over a range of different accelerations, and we can notice that for classical accelerations, \cref{eq:rel_u_EV} it is wrong in principle.

\section{Conclusions} \label{sec:conclusion}

In this paper we analysed two main articles \citep{mcculloch_modelling_2007,mcculloch_inertia_2013} describing Quantized Inertia. We found two major flaws on the derivation presented, and we propose some corrections to address the found issues. Such flaws, if they do not invalidate, at least will require a major rethinking of the whole theory. In our article, we did not address the ability of Quantized Inertia to match the observational data. 

We consider that speculative physics is fundamental for the constant progress of science: Quantized Inertia was often criticized because it does go against well-established principles such as the equivalence principle. We consider this should not be the criterion used to establish the validity of a theory:  history teaches us that many scientific breakthroughs, encountered, in the beginning, strong resistance from the scientific community because they were against existing principles. For this reason, it is of fundamental importance that any new iteration of quantized inertia should have a stronger mathematical derivation and, eventually, a strategy for a practical experimental verification.

\section*{Acknowledgements}
This work was supported by the research project \texttt{PN19060104}. We would like to thank our ATLAS group colleagues for the supportive working environment and in particular Prof. C\u{a}lin Alexa for his guidance and support during the writing of this article. We would like also to thanks the \citet{bokeh_2014} for the excellent tool used to create the plots of this paper and the reviewer of this article for his meaningful feedbacks and the intellectual integrity shown during the review process.




\bibliographystyle{mnras}
\bibliography{bibliography} 

%
%
%
%
%

\bsp	
\label{lastpage}
\end{document}